\documentclass[journal]{IEEEtran}
\usepackage{cite}

\ifCLASSINFOpdf
  \usepackage[pdftex]{graphicx}
\else
\fi
\usepackage{threeparttable}
\usepackage{multirow}
\usepackage{booktabs}
\usepackage{color}
\begin{document}
%
\title{Aspect-driven User Preference and News Representation Learning for News Recommendation}
%
%
%

\author{Rongyao~Wang,
        Wenpeng~Lu,
        Shoujin~Wang,
        Xueping~Peng,
        Hao~Wu,
        and~Qian~Zhang}
\maketitle

\begin{abstract}
Most of existing news recommender systems usually learn topic-level representations of users and news for recommendation, and neglect to learn more informative aspect-level features of users and news for more accurate recommendation. As a result, they achieve limited recommendation performance.
Aiming at addressing this deficiency, 
we propose a novel Aspect-driven News Recommender System (ANRS) built on aspect-level user preference and news representation learning.
Here, \textit{news aspect} is fine-grained semantic information expressed by a set of related words, which indicates specific aspects described by the news.
In ANRS, \textit{news aspect-level encoder} and \textit{user aspect-level encoder} are devised to learn the fine-grained aspect-level representations of user's preferences and news characteristics respectively, which are fed into \textit{click predictor} to judge the probability of the user clicking the candidate news.
Extensive experiments are done on the commonly used real-world dataset MIND, which demonstrate the superiority of our method compared with representative and state-of-the-art methods.
\end{abstract}


%
\IEEEpeerreviewmaketitle

\section{Introduction}
\label{intro}
%
%
%
%
\IEEEPARstart{N}{owadays}, with the rapid development of artificial intelligence, deep learning technology plays important roles in various tasks \cite{wang2019sequential,wang2021survey,ijcaiGraphReview,wang2018attention,wang2021hierarchical,lu2017motor,lu2021is}, which is also applied widely in news recommendation. 
For example, TANR \cite{wu2019neurala} is a neural topic-aware news recommender system, which trains the news encoder with an auxiliary topic classification task to  learn accurate news representations. 
NAML \cite{ijcai2019-536} leverages attentive mechanisms to learn user's preferences and news features from multi-view information including the title, category and body of news for news recommendation. 
GNewsRec \cite{hu2020graph} first builds a heterogeneous news-topic-user graph and then applies graph convolution networks (GCN) and long short-term memory (LSTM) on the graph to extract user's long- and short-term preferences for news recommendation. 
DAN \cite{zhu2019dan} captures user's latent preferences towards news with attention-based convolution neural networks (CNN) and recurrent neural networks (RNN) from clicked news for recommendation. 
Although these existing methods have achieved some progresses in news recommendation task, most of them attempt to model user's preferences and represent news features on coarse-grained topic level instead of fine-grained aspect level of news. 
However, aspect information is essential for news recommendation. 
For example, for a piece of news related with \textit{food}, its topic maybe \textit{food}, while it may include several detailed aspects, such as \textit{beef}, \textit{pasta} and \textit{tomato} \cite{he2017unsupervised}. It is obvious that the aspects are more accurate than the topic, which can provide more fine-grained information for news recommender systems.
However, the existing methods always neglect to model user's preferences and news features on the aspect level, which leads to the limited performance on news recommendation.

\begin{figure}
	\centering
	\includegraphics[width=\linewidth]{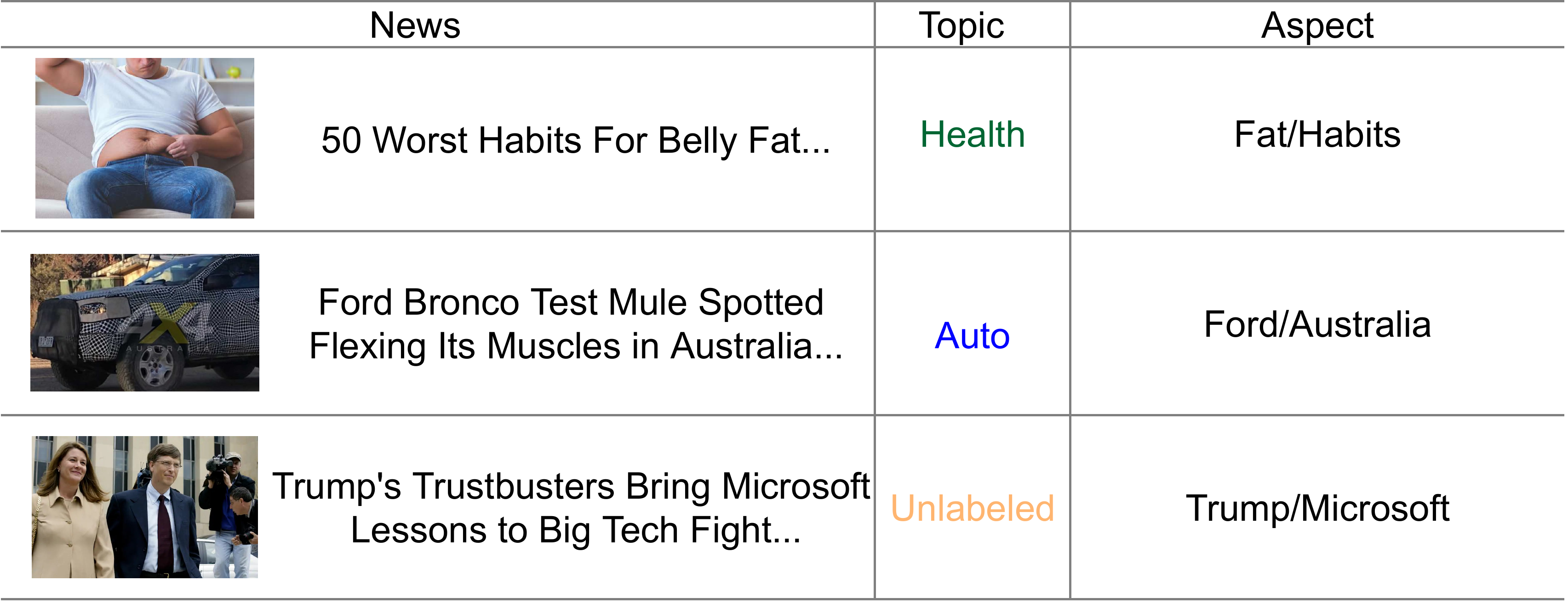}
	\caption{Examples of news and its corresponding topic and aspect}
	\label{fig1}
\end{figure}

According to our investigation on existing work, three observations for news recommendation are found.
First, though topic information is important for news, it is still coarser than aspect information, and hence it can not accurately describe news features. 
As illustrated in Fig. \ref{fig1}, the topic of the first news \textit{``50 Worst Habits For Belly Fat...''} is \textit{health}. 
Besides such topic information, the news could be described with fine-grained aspects, such as \textit{fat}, \textit{habits}, etc.  
Second, topic information can not comprehensively represent user preferences. It is necessary to utilize fine-grained aspect information to emphasize the specific preferences of a user. 
Taking the second news in Fig. \ref{fig1} as an example, a user may be interested in certain aspects, such as \textit{Ford} and \textit{Australia}. 
However, if the recommender system merely works on topic information, it would recommend all \textit{auto} news to the user, instead of \textit{Ford} and \textit{Australia} ones. 
The user's experience will be inevitably downgraded by the irrelevant news.  
Third, lots of news isn't annotated explicitly with topic information. 
Huge amounts of news is generated every day, and topic annotation is time-consuming, so it is impossible to manually annotate topic labels for all news. This limits the applicability of news recommender systems built on explicit topic information. In contrast, the aspect information of news can be extracted automatically. These observations mentioned above demonstrate the necessity of making news recommendation driven by the fine-grained aspect information. 

\par To this end, we propose a novel Aspect-driven News Recommender System (ANRS), which is built on aspect-level user preference and news representation learning. 
Specifically, ANRS model consists of three modules, i.e., \textit{user aspect-level encoder}, \textit{news aspect-level encoder} and \textit{click predictor}. 
\textit{News aspect-level encoder} is built on CNNs and attention networks to learn aspect-level news representations to capture the fine-grained semantic features.  
Similarly, \textit{User aspect-level encoder} is also built on CNNs and attention networks to accurately learn the representation of the user's aspect-level preference towards news.
More specifically, \textit{news aspect-level encoder} and \textit{user aspect-level encoder} are equipped with an \textit{aspect-level feature extractor} to generate aspect-level representations for news and users, which are further fed into \textit{click predictor} to calculate the probability of a given user to click on each candidate news and make the personalized recommendation for the user.
Extensive experiments are done the real-world dataset MIND, which demonstrate that our ANRS model is able to achieve superior performance compared with state-of-the-art competing models.
Our main contributions are concluded as follows.
\begin{itemize}
	\item We propose to make accurate news recommendation by modeling fine-grained aspect-level user preferences and news features. As far as we know, this is the first work for news recommendation driven by aspect-level information.     
	
	\item We devise a novel Aspect-driven News Recommender System (ANRS), which is built on aspect-level user preference and news representation learning. ANRS consists of  \textit{user aspect-level encoder}, \textit{news aspect-level encoder} and \textit{click predictor}. Particularly, \textit{aspect-level feature extractor} is elaborately designed to extract aspect information from news for better representing user's preferences and news features. 
	
	\item We conduct extensive experiments on the real-world news recommendation dataset MIND. The experimental results substantially verify the effectiveness of ANRS, which demonstrates better performance than state-of-the-art competing models. \footnote{The source code will be publicly available once the acceptance of this manuscript.
	}
	
\end{itemize}

\section{Related Work}
\label{related work}
We review two categories of representative and state-of-the-art recommendation methods: topic-based news recommendation and aspect-based product recommendation, which are the most relevant ones to our work.

Topic-based news recommender systems have been widely employed due to their strong capabilities to model topic information for news recommendation. 
The first topic-based news recommender system, TANR \cite{wu2019neurala}, employed CNN and attention networks, and combined with topic classification task to learn topic-aware news and user representations for news recommendation. 
Later, Wu et al. \cite{ijcai2019-536} proposed a multi-view learning model to generate news representations from their titles, bodies and topic categories for news recommendation. 
Then, Hu et al. \cite{hu2020graph} proposed a heterogeneous user-topic-news graph-based news recommender system, which learned user's long-term preference representations and news representations with graph neural network, and learned user's short-term preference representations with CNN combined with LSTM networks. 
Lee et al. \cite{lee2020news} devised a topic-enriched news recommender system with a knowledge graph, which exploited external knowledge and topical relatedness in news. 
Ge et al. \cite{ge2020graph} proposed a graph-enhanced news recommender system, which employed a modified transformer and graph attention network to encode news titles and topics so as to enhance news and user representations.
Although these topic-based news recommender systems have achieved great successes, they model user's preference and news features on the coarse-grained topic-level information, neglecting the fine-grained aspect-level information. This limits their further improvement on recommendation performance.

The definition of aspects originates from sentiment analysis task, which are extracted for analyzing which opinions have been expressed \cite{liu2012sentiment}. In a recommendation scenario, users often provide reviews to explain why they like or dislike products from different aspects, such as prices, quality, etc. The topic models, such as latent dirichlet allocation (LDA) \cite{blei2003latent}, probabilistic latent semantic analysis (PLSA) \cite{hofmann2001unsupervised}, are always employed to extract the aspects. To accurately capture the aspect information is crucial for recommender systems to recommend next items for users \cite{chin2018anr}.
In recent years, the aspect-based product recommender systems focus on extracting aspects from textual reviews of products.
ALFM \cite{cheng2018aspect} was an aspect-aware topic model, which employed a topic model to extract the topic representations of aspects and aspect-aware latent model to evaluate the overall user-product rating. 
Bauman et al. \cite{bauman2017aspect} proposed aspect-based recommendation with user reviews, which recommended products together with important aspects to users according to the most valuable aspects of the user's potential experience.
Hou et al. \cite{hou2019explainable} devised an explainable recommender system with aspect information, which utilized aspect-level information to make quantitative fine-grained explanations for recommendation.
ANR \cite{chin2018anr} was an end-to-end aspect-based neural recommendation model, which learned aspect-based representations for both users and products with an attention-based component.
Luo et al. \cite{luo2019hybrid} proposed an aspect-based explainable recommender system, which addressed some problems including dynamic of explanation, personal preference learning and the granularity of explanations.
Much progress has been achieved for aspect-based product recommendation, demonstrating the significance of aspect-level information for recommendation. 

Although aspect-level information have been widely applied in product recommendation task, there is no existing work in news recommendation that is driven by aspect-level information. According to the former work on aspect-based product recommender systems, it is obvious that aspect information is more fine-grained than topic information, which can express more accurate semantic information and is beneficial to improve the recommendation performance. In a news recommendation scenario, users may also like or dislike the news from various aspects, such as \textit{politicians}, \textit{epidemic situation}, \textit{country}, etc. Considering the great success of aspect information in product recommendation, we introduce it into news recommendation and propose aspect-driven user preference and news representation learning for news recommendation. 

\section{Preliminaries}
\subsection{Problem Statement}
Given a user \( u \), a set of news browsed by \( u \) is defined as \({D^u} = \left\{ {{d_1},{d_{2}}, \cdots,{d_m}} \right\} \).   
Given a piece of candidate news \( {c} \) with a binary label \( {y} \in \left\{ {0,1} \right\} \) which indicates whether \( u \) will click \( {c} \).
ANRS is trained as a prediction model which learns to predict the probability that \( u \) would like to click \( {c} \) according to the browsed news $D^u$. The probabilities are used to rank all candidate news.
Once the parameters of the model are trained, ANRS is able to perform personal news recommendation based on the ranking of candidate news.

\subsection{Definition of News Aspect}
News aspect information is more fine-grained than topic-level information, which is beneficial to the accurate news recommendation. 
Inspired by the work of He et al. \cite{he2017unsupervised}, 
we define \textit{news aspect} as a group of representative words in a piece of news.

\begin{table}
	\caption{Examples of news aspects.}\label{tab1}
	\centering
	\begin{tabular}{|c|c|}
		\hline
		\textbf{Aspects}&\textbf{Representative words}\\
		\hline
		Festival&celebrate, Christmas, parade, holiday, turkey\\
		\hline
		Murder&detect, evidence, arrest, charged, shot\\
		\hline
		Weather&winter, sunny, storm, forecast, rain\\
		\hline
	\end{tabular}
\end{table}

Some examples of news aspects are provided in Table \ref{tab1}. We can find that the aspect of news is composed of a set of words in news, which is different from news topic.
For example, the second news in Fig. \ref{fig1} has been annotated manually with topic label \textit{Auto} and sub-topic label \textit{Autonews}. Both of labels belong to the same category, which are coarse-grained topic information. 
However, our model can capture different fine-grained news aspects, such as \textit{Ford} and \textit{Australia}.
These fine-grained aspect-level information can reflect more accurate features of news, which are crucial for the performance improvement of news recommender systems.

\section{ Aspect-driven News Recommender System}
\begin{figure*}[t]
	\centering
	\includegraphics[width=1.0\textwidth]{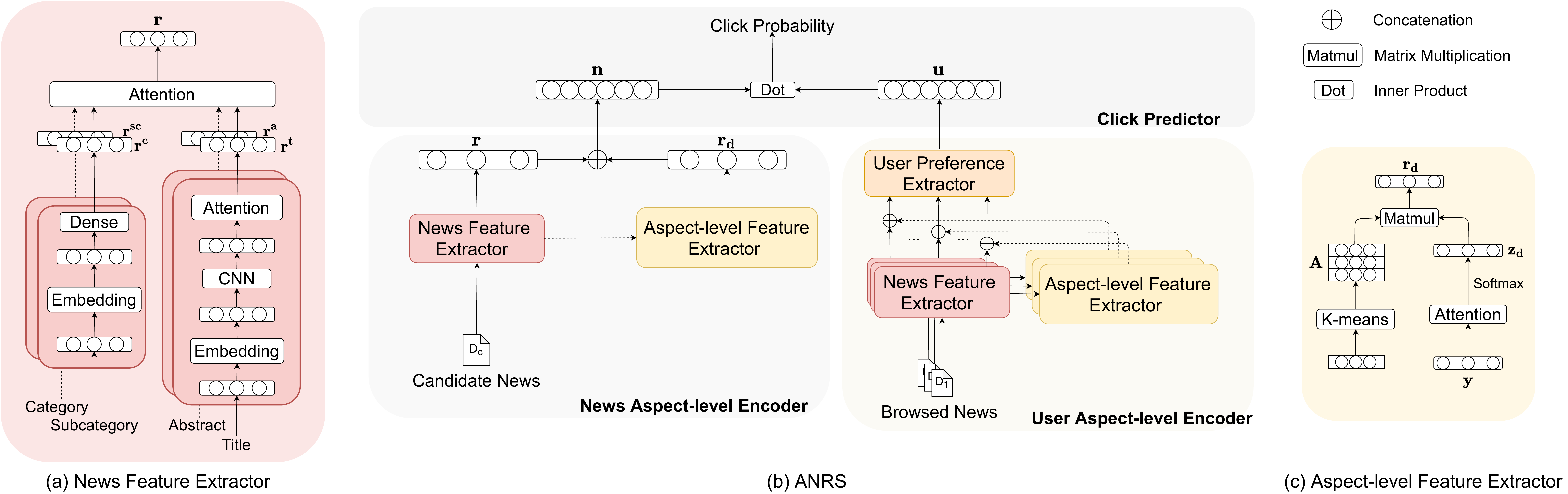}
	\caption{Model architecture of ANRS, consisting of three modules: \textit{news aspect-level encoder}, \textit{user aspect-level encoder} and \textit{click predictor}. Specifically, \textit{news aspect-level encoder} includes two core sub-modules: \textit{news feature extractor} and \textit{aspect-level feature extractor}. Moreover, \textit{user aspect-level encoder} contains three core sub-modules: \textit{news feature extractor},  \textit{aspect-level feature extractor} and \textit{user preference extractor}. 
	}
	\label{fig:framework}
\end{figure*}

In this section, we introduce our model illustrated in Fig. \ref{fig:framework}.
The architecture of ANRS contains three core modules: \textit{news aspect-level encoder}, \textit{user aspect-level encoder} and \textit{click predictor}.
\textit{News aspect-level encoder} is utilized to encode news aspect-level information, which contains \textit{news feature extractor} and \textit{aspect-level feature extractor}.
\textit{User aspect-level encoder} is employed to encode user aspect-level preference, which consists of \textit{user preference extractor}, \textit{news feature extractor} and \textit{aspect-level feature extractor}.
\textit{Click predictor} is applied to calculate the clicked probability of candidate news.
We describe more details of these modules in this section.

\subsection{News Aspect-level Encoder}
\label{sec:news encoder}
In order to obtain news feature embeddings and aspect-level feature embeddings, we design \textit{news aspect-level encoder}, which consists of two core sub-modules:  \textit{news feature extractor} and  \textit{aspect-level feature extractor}, as shown in Fig. \ref{fig:framework}.
Given a piece of news, first, \textit{news feature extractor} and \textit{aspect-level feature extractor} are employed to encode it to generate the initial news embedding \( {\bf{r}} \) and the aspect-specific embedding \({{\bf{r}}_{d}} \) respectively. Then, 
both \( {\bf{r}} \) and \({{\bf{r}}_{d}} \) are concatenated together to obtain the final news representation \( {{\bf{n}}} \), described as follows:
\begin{equation}
	{{\bf{n}}} = \left[ {{{\bf{r}}};{{\bf{r}}_{d}}} \right] ,
	\label{e2:concate}
\end{equation} 
where the operator $[ ; ]$ means the concatenation operation.

\subsubsection{\textbf{News Feature Extractor}}
Aiming to fully extract information contained in three sections of each news, i.e., title, category and abstract, inspired by the work of Wu et al. \cite{ijcai2019-536}, we design four components to encode them  into latent representations simultaneously, i.e., \textit{title learning component}, \textit{abstract learning component}, \textit{category learning component} and \textit{attention component}, as illustrated in Fig. \ref{fig:framework} (a).

\textbf{\textit{Title learning component.}} There are three layers in this component. The first layer is to transform the words in news titles into word embeddings.  
We define a news title as ${T} = \left[ {{t_1},{t_2}, \cdots, {t_N}} \right]$, where $N$ refers to the length of ${T}$, \(t\) denotes a word in ${T}$. 
This layer transforms \(T\) into the embedding representation \({{\bf{E}}^n} = \left[ {{{\bf{e}}_1}{\bf{,}}{{\bf{e}}_2}{\bf{,}} \cdots{\bf{,}}{{\bf{e}}_N}} \right]\) according to the word embedding matrix \({{\bf{W}}} \in {{\bf{R}}^{V \times D}}\), where \(V\) and \(D\) denote the size of vocabulary and the dimension of word embedding, respectively.

CNN is widely applied to capture text features and demonstrates better performance in news recommendation.
The previous works of FIM \cite{wang2020fine}, NPA \cite{wu2019npa} and NAML \cite{ijcai2019-536} employ CNN to learn local features of news.
Hence, we utilize CNN to capture the deeper features in the second layer of \textit{title learning component}.
The representation of the \textit{i}-th position is described as Equ. (\ref{e3:CNN}).
\begin{equation}
	{{\bf{c}}_i} = f \left( {{{\bf{C}}_w} * {{\bf{e}}_{\left({i - k} \right):\left( {i + k} \right)}} + {{\bf{b}}_w}} \right),
	\label{e3:CNN}
\end{equation}
where $f$ is ReLU, $*$ refers to the convolution operator, \({{\bf{e}}_{\left( {i - k} \right):\left( {i + k} \right)}}\) denotes the concatenation of word embeddings from the position \(\left( {i - k} \right)\) to \(\left( {i + k} \right)\), 
\({{\bf{C}}_w}\) is the kernel of CNN filters, \({{\bf{b}}_w}\) is the bias.  
After the process of this layer, we can obtain a sequence of contextual word embeddings, i.e., \(\left[ {{{\bf{c}}_1},{{\bf{c}}_2},\cdots,{{\bf{c}}_N}} \right]\).

The importance of each word in a piece of text is different. It is crucial to find the important features in news representations \cite{zhu2019dan}. Therefore,
in the third layer of \textit{title learning component}, we utilize a word-level attention network \cite{wu2019b} to enhance the significant features and obtain informative representations of news titles, as described in the following equations.

\begin{equation}
	a_i = {\bf{q}}^\top \cdot \sigma \left( {{{\bf{V}}} \cdot {\bf{c}}_i + {{\bf{v}}}} \right) ,
	\label{e4:attention1}
\end{equation}
\begin{equation}
	\alpha_i = \frac{{\exp \left( {a _i} \right)}}{{\sum\nolimits_{j = 1}^N {\exp \left( {a _j} \right)} }} ,
	\label{e5:attention2}
\end{equation}
\begin{equation}
	{{\bf{r}}^t} = \sum\limits_{j = 1}^N {{{\alpha_j}}} {{\bf{c}}_j} ,
	\label{e6:sum of attention}
\end{equation}
where the symbol $\sigma$ is tanh, \({\bf{q}}\), \({{\bf{V}}}\) and \({{\bf{v}}}\) are parameters learnt by the training process, ${\bf{c}}_j$ is $j$-th word embedding of the title, 
\( {{\bf{r}}^t} \) is the final representation of news title.

\textbf{\textit{Abstract learning component.}} 
The learning procedure for news abstract is almost same with that for news title. Therefore, \textit{abstract learning component} is similar with the former one, i.e., \textit{title learning component}, whose output is the representation of news abstract, marked with ${\bf{r}}^a $.

\textbf{\textit{Category learning component.}} 
Some news is annotated manually with category and subcategory labels, which also represent key features of the news.
In order to learn news category and sub-category information, we devise \textit{category learning component}, which converts two kinds of categories into low-dimensional dense representations and generates corresponding representations, described as follows:

\begin{equation}
	{{\bf{r}}^c} = f \left( {{{\bf{V}}_c} \cdot {{\bf{e}}^c} + {{\bf{v}}_c}} \right) ,
	\label{e7:rc}
\end{equation}
\begin{equation}
	{{\bf{r}}^{sc}} = f \left( {{{\bf{V}}_{sc}} \cdot {{\bf{e}}^{sc}} + {{\bf{v}}_{sc}}} \right) ,
	\label{e8:rsc}
\end{equation}
where $f$ is ReLU, \( {{\bf{e}}^c} \) and \( {{\bf{e}}^{sc}} \) refer to the category and sub-category embedding,  \( {{\bf{V}}_c} \), \( {{\bf{v}}_c} \), \( {{\bf{V}}_{sc}} \) and \( {{\bf{v}}_{sc}} \) denote the parameters in dense layers, ${\bf{r}}^c$ and ${\bf{r}}^{sc}$ are the representations of news category and sub-category, respectively.

\textbf{\textit{Attention component.}} 
Following the previous work of Wu et al. \cite{ijcai2019-536}, we apply an attention network to learn the weights of various information in news and then build the final news representation.
The method for evaluating the attention weight of title, i.e., $\alpha_{t}$, is described as follows:

\begin{equation}
	a_t = {\bf{q}}_t^\top \cdot \sigma \left( {{{\bf{V}}_t} \cdot {{\bf{r}}^t} + {{\bf{v}}_t}} \right) ,
	\label{a_t}
\end{equation}
\begin{equation}	{\alpha _t} = \frac{{\exp \left( {{a_t}} \right)}}{{\exp \left( {{a_t}} \right) + \exp \left( {{a_a}} \right) + \exp \left( {{a_c}} \right) + \exp \left( {{a_{sc}}} \right)}} ,
	\label{alpha_t}
\end{equation}
where $\sigma$ is tanh, \({{\bf{q}}_t}\), \({{\bf{V}}_t}\) and \({{\bf{v}}_t}\) are the learnable parameters. For $a_a$, $a_c$ and $a_{sc}$ of news abstract, category and sub-category, they are obtained with the similar equations as Equ. (\ref{a_t}).
The attention weights of abstract, category and subcategory are evaluated with the similar procedure as the weight of title, i.e., Equ. (\ref{alpha_t}), which are denoted as $\alpha_a$, $\alpha_c$ and $\alpha_{sc}$ respectively.

Finally, the news feature embedding \( {\bf{r}} \) is built by concatenating news title embedding ${\bf{r}}^t$, news abstract embedding ${\bf{r}}^a$, news category embedding ${\bf{r}}^c$, and news sub-category embedding ${\bf{r}}^{sc}$ according to their different attention weights, described with the following equation:
\begin{equation}
	{\bf{r}} = \left[ {{\alpha_t}{{\bf{r}}^t};{\alpha_a}{{\bf{r}}^a};{\alpha_c}{{\bf{r}}^c};{\alpha_{sc}}{{\bf{r}}^{sc}}} \right] .
	\label{r}
\end{equation} 
\subsubsection{\textbf{Aspect-level Feature Extractor}}
As shown in Fig. \ref{fig:framework} (c), inspired by the previous work of He et al. \cite{he2017unsupervised}, we devise the aspect-level feature extractor, which consists of \textit{attention-based generation of news embedding} and \textit{aspect-based reconstruction of news embedding}.
The process of aspect-level feature extractor is similar to autoencoders, where we employ attention mechanism to extract key aspect words in each news and generate the news embedding with the weighted word embeddings, then reconstruct each news by the combinations of aspect embeddings. 

\textbf{\textit{Attention-based generation of news embedding.}} 
In order to capture the important aspect words in each news, we utilize an attention mechanism to encode the news, described as the following equations:

\begin{equation}
	{{\bf{y}}} = \frac{1}{N}\sum\limits_{i = 1}^N {{{\bf{e}}_i}} ,
	\label{e10:z}
\end{equation}

\begin{equation}
	{h_i} = {\bf{e}}_{i}^ \top  \cdot {\bf{H}} \cdot {{\bf{y}}} ,
	\label{e11:y}
\end{equation}

\begin{equation}
	\label{e12:a'}
	{\alpha'_i} = \frac{{\exp \left( {{h_i}} \right)}}{{\sum\nolimits_{j = 1}^N {\exp \left( {{h_j}} \right)} }},
\end{equation}

\begin{equation}
	{{\bf{z}}_{d}} = \sum\limits_{i = 1}^N {\alpha'_i} {{\bf{e}}_{i}},
	\label{e13:z}
\end{equation}
where ${\bf{e}}_i$ is the embeddings of the $i$-th word in news content, including its title, abstract and category. $\bf{y}$ is the average of the word embeddings, which can be viewed as a simple global news representation. $\bf{H}$ is a matrix for mapping $\bf{y}$ and ${\bf{e}}_i$, which is learned in the training process. $\alpha'_i$ is the weight, which can be viewed as the probability of the $i$-th word is the right aspect word to describe the main aspect information of the news. ${\bf{z}}_{d}$ is the attention-based news embedding according to the weighted probability of each possible aspect word. 

\textbf{\textit{Aspect-based reconstruction of news embedding.}} 
In order to assure the quality of aspect extraction, inspired by autoencoders, we reconstruct each news through a liner combination of the extracted aspect embeddings, described as follows:

\begin{equation}
	{{\bf{p}}} = softmax\left( {{{\bf{W}}_p} \cdot {{\bf{z}}_{d}} + {\bf{b}}} \right) ,
	\label{e14:p}
\end{equation}
\begin{equation}
	{{\bf{r}}_{d}} = {{\bf{A}}^ \top } \cdot {{\bf{p}}} ,
	\label{e15:r'}
\end{equation}
where ${\bf{W}}_p$ is weighted matrix parameter,
$\bf{b}$ is the bias vector. 
$\bf{p}$ can be viewed as the weight vector over all aspect embeddings.
$\bf{A}$ is the aspect embedding matrix, which is initialized with the $k$-means centroids of the news embeddings. 
${\bf{r}}_{d}$ is the news embedding reconstructed with aspect words.

\subsection{User Aspect-level Encoder}

As shown in Fig. \ref{fig:framework} (b), aiming to model the user preferences accurately, we devise \textit{user aspect-level encoder}, which consists of \textit{news feature extractor}, \textit{aspect-level feature extractor} and \textit{user preference extractor}.
Specifically, \textit{news feature extractor} and \textit{aspect-level feature extractor} learn the general news representation \( {\bf{k}} \) and aspect-specific news representation \({{{\bf{k}}}_{d}}\) respectively. Both extractors are same with the corresponding modules of news aspect-level encoder in Section \ref{sec:news encoder}.
The concatenation of ${\bf{k}}$ and ${\bf{k}}_{d}$ is the final representation of the browsed news, as described below.
\begin{equation}
	{{\bf{n'}}} = \left[{{\bf{k}};{{{\bf{k}}}_{d}}} \right].
\end{equation}

In order to model the user representations from their browsed news, we further devise \textit{user preference extractor}. It employs a sentence-level attention mechanism to choose important news and learn more accurate user representations as below:

\begin{equation}
	a_i^n = {\bf{q}}_n^\top \cdot \sigma \left( {{{\bf{V}}_n} \cdot {{\bf{n}}'_i} + {{\bf{v}}_n}} \right) ,
	\label{e17:a}
\end{equation}
\begin{equation}
	\alpha _i^n = \frac{{\exp \left( {a_i^n} \right)}}{{\sum\nolimits_{j = 1}^M {\exp \left( {a_j^n} \right)} }} ,
	\label{e15:an}
\end{equation}
\begin{equation}
	{\bf{u}} = \sum\limits_{i = 1}^M {\alpha _i^n{{\bf{n}}'_i}} ,
\end{equation}
where $\sigma$ is tanh operation, \( {{\bf{n}}'_i} \) is the representation of the \textit{i}-th browsed news by user $u$, \({{\bf{q}}_n}\), \({{\bf{V}}_n}\) and \({{\bf{v}}_n}\) are the learnable parameters, \(M\) denotes the number of browsed news of the user $u$, $\alpha_i^n$ refers to the attention weight of the $i$-th browsed news. 
The aspect-level representation $\bf{u}$ of the user $u$ is the weighted summation of the browsed news representations.

\subsection{Click Predictor}

Once obtaining the aspect-level news representation ${\bf{n}}$ and user representation ${\bf{u}}$, \textit{click predictor} is devised to calculate the probability of a user clicking the candidate news.
Following the work of Okura et al. \cite{okura2017embedding}, we obtain the click score \(\hat y\) by calculating the inner product of these representations between the user $u$ and the news $n$:
\begin{equation}
	\hat y = {\bf{u}}^ \top  \cdot {{\bf{n}}} .
	\label{e20:y}
\end{equation} 
\subsection{Optimization and Training}
Following the work of Wu et al. \cite{wu2019neurala}, given a user $u$, we define his browsed news as positive samples and randomly select some negative samples according to the negative sample ratio.  
The click probability of a given positive sample $y_i^+$ w.r.t. $G$ sampled negative samples \(\left[ {\hat y_1^-, \hat y_{\rm{2}}^ -, \cdots,\hat y_G^ - } \right]\) is calculated as:

\begin{equation}
	{p_i} = \frac{{\exp \left( {\hat y_i^ + } \right)}}{{\exp \left( {\hat y_i^ + } \right) + \sum\limits_{j = 1}^G {\exp \left( {\hat y_{i,j}^ - } \right)} }} ,
	\label{e21:p}
\end{equation}
where ${\hat y_{i,j}^ - }$ is the click score of the $j$-th negative sample in the same session with the $i$-th positive sample.
The negative log-likelihood of all positive samples is the loss function for news recommendation, described as:

\begin{equation}
	U(\theta) =  - \sum\nolimits_{i \in {\cal P}} {\log \left( {{p_i}} \right)} ,
	\label{e22:lossnr}
\end{equation}
where \({\cal P}\) is the set of positive samples.

In order to ensure the quality of aspect-level representation extracted by the model and ensure the diversity of aspect embeddings, motivated by the work of He et al. \cite{he2017unsupervised}, we devise the loss function to constrain aspect-level information, as follows:

\begin{equation}
	J\left( \theta  \right) = \sum\nolimits_{d \in \cal S} {\sum\limits_{j = 1}^G {\max \left( {0,1 - {{\bf{r}}_{d}}{{\bf{z}}_{d}} + {{\bf{r}}_{d}}{{\bf{n}}_j}} \right)} } ,
	\label{e23:J}
\end{equation}
where \(\cal S\) is the training dataset,
\({\bf{r}}_{d}\) and \({\bf{z}}_{d}\) is generated by \textit{aspect-level feature extractor},
\( {{\bf{n}}_j} \) is the representation of a negative sample.
\begin{equation}
	F\left( \theta  \right) = \left\| {{{\bf{A}}_n} \cdot {\bf{A}}_n^ \top  - {\bf{I}}} \right\| ,
	\label{E24:F}
\end{equation}
where \(F\) is regularization term, which is able to ensure the diversity and uniqueness of aspect embedding.
\({{\bf{A}}_n}\) is \({\bf{A}}\) with each row normalized to length 1. 
\({\bf{I}}\) is the identity matrix.

The final loss \(Los{s_{final}}\) is obtained by merging the above three losses:
\begin{equation}
	Los{s_{final}} = U(\theta) + J\left( \theta  \right) + \lambda F\left( \theta  \right) ,
	\label{e25:lossfinal}
\end{equation}
where \(\lambda\) can control the weight of the regularization term.

\section{Experiment and Evaluation}
In this section, we carry out extensive experiments on a public real-world dataset and compare ANRS with the popular baselines. 
Besides, we conduct some analysis experiments to investigate the influence of hyperparameters in ANRS.
Finally, a case study is implemented.
\subsection{Experimental Setup}
\subsubsection*{\textbf{Datasets}}
\begin{table}[htbp]
	\caption{The statistics on the MIND dataset.}
	\label{tab2}
	\centering
	\begin{tabular}{lll}
		\hline
		\textbf{Stats.} & \textbf{MIND-small} & \textbf{MIND-large}\\
		\hline
		News & 65,238 & 161,013\\
		Topics & 18 & 20 \\
		Users & 94,057 &  1,000,000\\ 
		Clicks & 347,727 & 24,155,470 \\ 
		\hline
	\end{tabular}	
\end{table}
We experimented ANRS on a commonly used real-world dataset MIND released on ACL 2020 \cite{wu2020mind}. It is collected from anonymous behavior logs of Microsoft News website, which contains two versions of different types including MIND-small and MIND-large.

Specifically, MIND-large has 24,155,470 logs of 1,000,000 users and 161,013 news from 20 topics, which is divided into training, validation and test sets. 
Moreover, MIND-small includes 347,727 logs of 94,057 users and 65,238 news from 18 topics, which is divided into training and validation sets. 
For each piece of news, MIND provides its title, category, abstract and entity. However, the detailed content of news is missing.
Its statistics is shown in Table \ref{tab2}.
Due to the limitation of Microsoft's license, we fail to access the labels of test set, and thus it cannot be used to test our model. To solve the problem, we take the released validation set as our test set, and split \({\rm{10\% }}\) samples from training set as the new validation set. 
\subsubsection*{\textbf{Parameter Settings}}
We utilize the pretrained Glove embedding \cite{pennington2014glove} to initialize word embedding and set the dimension to 300. 
For all modules, the filters of CNNs are set to 400 and window sizes are set to 5. For training, we set the negative sampling ratio to 6 and the batch size to 256. 
To avoid overfitting problem, the dropout is set to 0.2.
In order to obtain the best performance, the number of clusters is set to 40 for $k$-means clustering, which is the number of aspects. 
For evaluating the performance, we adopt three popular metrics, including AUC, MRR and nDCG \cite{hu2018interpretable,wu2019neurala}.
\subsubsection*{\textbf{Baselines}}
We compare ANRS with several representative and/or state-of-the-art baselines, including latent factor models and neural network models: 
\begin{itemize}
	\item \textbf{BiasMF} \cite{koren2009matrix}, a matrix factorization model for recommendation task.
	\item \textbf{FM} \cite{rendle2010factorization}, another non-linear model based on matrix factorization for recommendation task.
	\item \textbf{CNN} \cite{kim2014convolutional}, a classical convolution neural network, which encodes the word sequences of news titles and applies max pooling to capture features.
	\item \textbf{DKN} \cite{wang2018dkn}, a deep learning based news recommender system, which utilizes CNN and attention mechanisms to obtain user and news representations, and utilizes knowledge graph to improve the effectiveness of recommendation.
	\item  \textbf{Hi-Fi Ark} \cite{Liu2019Hi}, another deep learning based news recommender system, which proposes a user representation framework. It aggregates user history into archives to learn user representations.
	\item \textbf{TANR} \cite{wu2019neurala}, a state-of-the-art news recommender system, which generates topic-aware news representations with the help of topic category labels.
	\item \textbf{NRMS} \cite{wu2019neural}, an attention-based news recommender system which utilizes multi-head self-attention and additive attention networks to learn news and user representations. 
	\item \textbf{LSTUR} \cite{An2019LSTUR}, a sequence-based user model for news recommendation modeling the long and short term user representations via a GRU network based on clicked news.
\end{itemize}

In order to further verify the effectiveness of aspect information, we implement the simplified ablation version of ANRS, marked as \textbf{ANRS$^{-a}$}, which removes all aspect-level feature extractors from the standard ANRS model.
\subsection{Performance Evaluation}
In order to answer the following four questions, we conduct extensive experiments:

Q1: How does our model perform compared with the baselines and what are the improvements?

Q2: How does aspect information of news affect the recommendation performance?

Q3: How does the pre-defined number of aspects affect the performance of our model? 

Q4: How does the different input data affect the performance of our model? 
\begin{table*}[htbp]
	\centering
	\caption{Performance Comparison on MIND dataset.}
	\begin{tabular}{c|cccc|cccc}
		\toprule
		\multirow{2}[4]{*}{Model} & \multicolumn{4}{c|}{MIND-small} & \multicolumn{4}{c}{MIND-large} \\
		\cmidrule{2-9}          & AUC   & MRR   & nDCG@5 & nDCG@10 & AUC   & MRR   & nDCG@5 & nDCG@10 \\
		\midrule
		baisMF & 0.5108  & 0.2258  & 0.2318  & 0.2952  & 0.5111  & 0.2257  & 0.2346  & 0.2963  \\
		FM    & 0.5004  & 0.2110  & 0.2110  & 0.2771  & 0.5084  & 0.2208  & 0.2312  & 0.2914  \\
		CNN   & 0.5073  & 0.2251  & 0.2316  & 0.2943  & 0.5071  & 0.2163  & 0.2202  & 0.2844  \\
		DKN   & 0.5726  & 0.2339  & 0.2418  & 0.3033  & 0.6329  & 0.2902  & 0.3163  & 0.3930  \\
		LSTUR & 0.6021  & 0.2659  & 0.2873  & 0.3529  & 0.5633  & 0.2454  & 0.2583  & 0.3252  \\
		NRMS  & 0.6391  & 0.3017  & 0.3282  & 0.3937  & \underline{0.6701}   & \underline{0.3185}  & \underline{0.3534}  & \underline{0.4175}   \\
		HiFi-Ark & 0.6403  & 0.2996  & 0.3272  & 0.3925  & 0.6394  & 0.2969  & 0.3221  & 0.3888  \\
		TANR  & \underline{0.6455} & \underline{0.3107} & \underline{0.3367} & \underline{0.4017}  & 0.6611  & 0.3148  & 0.3467  & 0.4114  \\
		\midrule
		ANRS$^{-a}$  & 0.6506  & 0.3136  & 0.3431  & 0.4076  & 0.6761  & 0.3185  & 0.3534  & 0.4175  \\
		ANRS  & \textbf{0.6673 } & \textbf{0.3235 } & \textbf{0.3569 } & \textbf{0.4183 } & \textbf{0.6826 } & \textbf{0.3350 } & \textbf{0.3722 } & \textbf{0.4343 } \\
		\midrule
		Improv.$^1$ & 3.38\% & 4.12\% & 6.00\% & 4.13\% & 1.87\% & 5.18\% & 5.32\% & 4.02\% \\
		\bottomrule
	\end{tabular}%
	\begin{tablenotes}
		\centering
		\footnotesize
		\item[1] $^{1}$ \small{Improvement achieved by ANRS over the best-performing baseline (TANR and NRMS) respectively.}
	\end{tablenotes}
	\label{tab3}%
\end{table*}%
\subsubsection*{Result1: Comparison with Baselines}
To demonstrate the effectiveness of ANRS, we compare the recommendation accuracy of our model with those baselines.
Table 3 reports the results of AUC, MRR, nDCG@5 and nDCG@10.
The first two methods, i.e., BiasMF and FM, are traditional latent factor models, which can not capture complex and deep representations effectively.
Obviously, they are beaten heavily by the others. 
Second, the single neural network, i.e., CNN only achieves similar performance with the latent factor models, which means that the basic neural model is unable to effectively capture the features in news contents.
Third, the deep neural network with news categories, i.e., DKN, LSTUR, Hi-Fi Ark and TANR, which achieve significant improvement than the latent factor models. 
Though the four models are superior to the others, they fail to capture fine-grained aspect information in news contents, which still limits their performance.
Fourth, the attention-based method, i.e., NRMS firstly applies the multi-head self-attention to model news representations for news recommendation.
This also ignores fine-grained aspect information to encode user representations, although it has the suboptimal performance compared with ANRS in terms of all metrics on the MIND-large dataset.  
Our proposed ANRS model not only captures various news information such as title and category, but also learns fine-grained aspect-level information.
As shown in Table \ref{tab3}, ANRS achieves the best performance on all metrics.
Especially, ANRS achieves a better performance than TANR, which is a state-of-the-art topic-based news recommender system. This demonstrates that the fine-grained aspect information is more powerful than the traditional topic information for news recommendation.
To be specific, in terms of AUC, MRR and nDCG, ANRS demonstrates significant improvement over the best-performing baseline, i.e., TANR, NRMS.

\subsubsection*{Result 2: Effectiveness of Aspect Information}
To demonstrate the effectiveness of aspect information, we remove the module \textit{aspect-level feature extractor} from ANRS to build an ablation variant ANRS$^{-a}$, and conduct experiments on the same dataset. 
As illustrated in Table \ref{tab3}, ANRS achieves better performance than ANRS$^{-a}$ in all metrics.
Specifically, in terms of AUC, ANRS is at least 1\% higher than that of ANRS$^{-a}$.
This may because the latter ANRS$^{-a}$ removes all aspect-related components and ignores all aspect-level information, which leads that ANRS$^{-a}$ achieves worse performance. 
It is obvious that the aspect-level information plays a great role in improving the performance of news recommendation.

\subsubsection*{Result 3: Effectiveness of the Number of Aspects}
\begin{figure}[htbp]
	\centering
	\includegraphics[width=6.5cm]{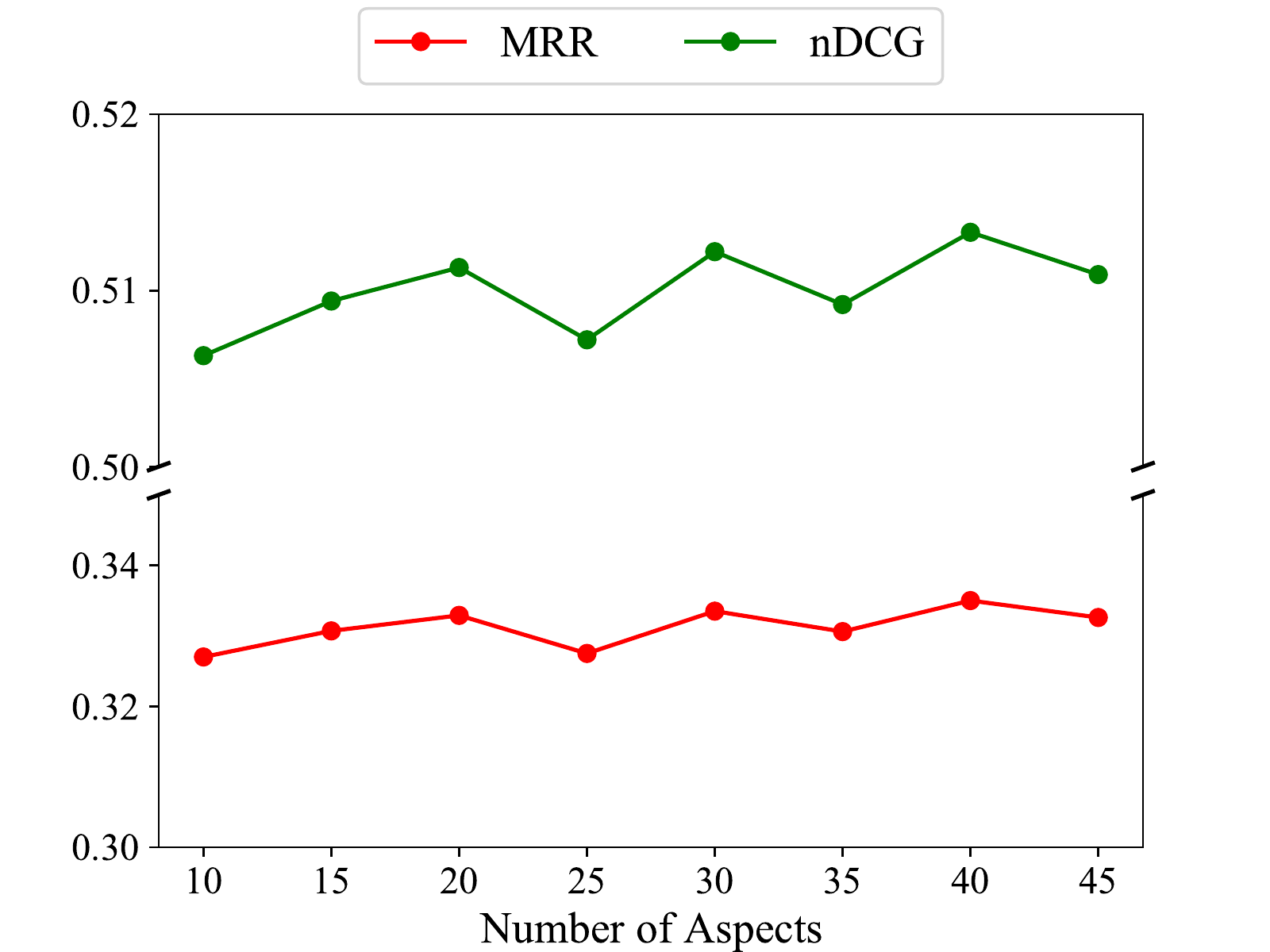}
	\caption{Performance of ANRS under different number of aspects}
	\label{aspects}
\end{figure}
To answer question Q3, we set the number of aspects from 10 to 45 and show their performances in Fig. \ref{aspects}. 
The aspect matrix in \textit{aspect-level feature extractor} is initialized with $k$-means centriods of the news embeddings, which determines the number of aspects.
When the number of aspects is set to 40, our model can achieve the best performance.
The possible reasons are two-fold. On the one hand, when the number is smaller than 40, the model is unable to capture enough aspect features. On the other hand, when the number is larger than 40, the model is easy to capture and induce noise aspect features.
Therefore, according to Fig. \ref{aspects}, we adopt 40 as the number of aspects in our model. 

\subsubsection*{Result 4: Influence of Different Input Data}
\begin{figure}[htbp]
	\centering
	\includegraphics[width=6.5cm]{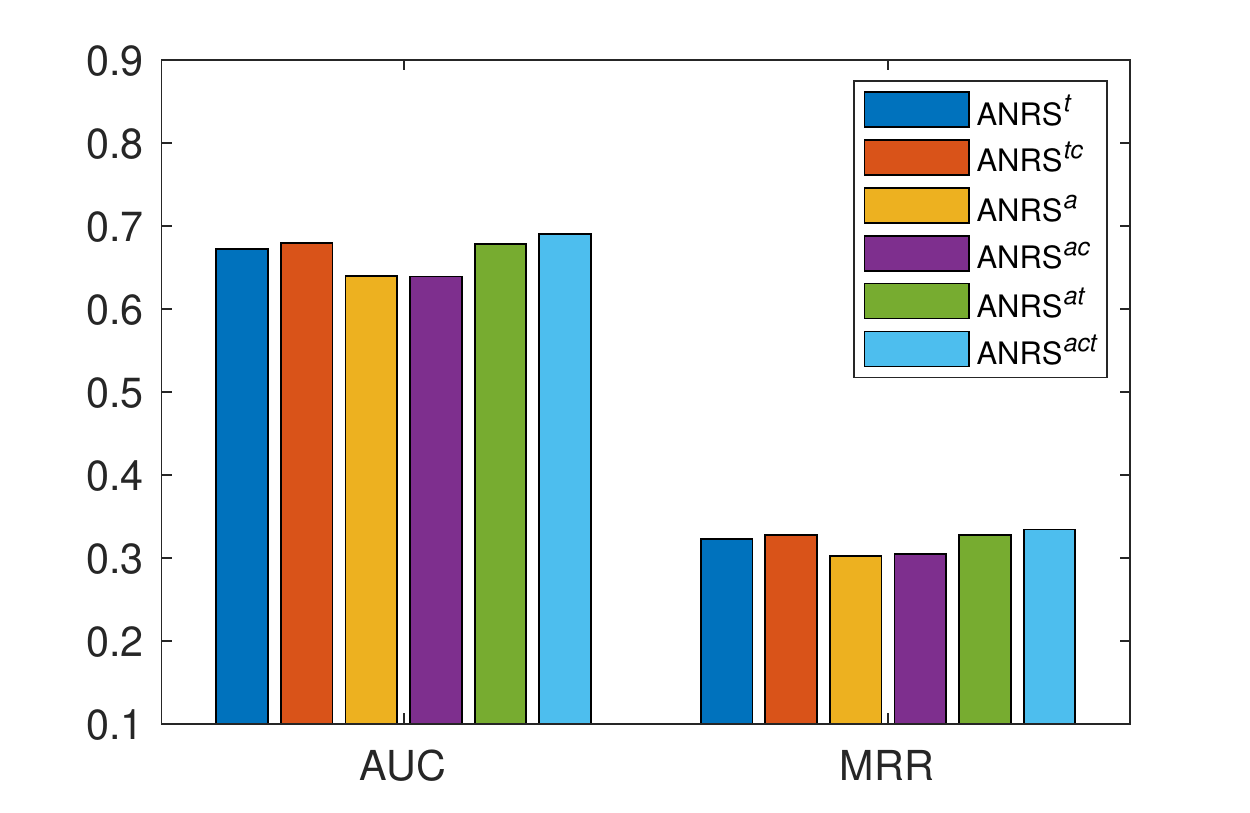}
	\caption{Performance of ANRS with different input data}
	\label{inputs}
\end{figure}
In order to answer question Q4, we feed different data in news into our model and compare the performances.
According to the difference of input data, the standard ANRS are transformed into six variants, that is, ANRS$^{t}$, ANRS$^{tc}$, ANRS$^{a}$, ANRS$^{ac}$, ANRS$^{at}$ and ANRS$^{act}$. The superscript $t$, $c$ and $a$ mean titles, categories and abstracts of news are fed into the corresponding model. 
As illustrated in Fig. \ref{inputs}, we can find that ANRS$^{act}$, i.e., the standard ANRS model, achieves the best performance. The variant is fed with all data, which is intuitive to achieve the best performance. 
In addtion, ANRS$^{t}$, ANRS$^{tc}$ show the better performance than ANRS$^{a}$, ANRS$^{ac}$ respectively.
This may because the title is more efficient data than the abstract in news recommendation.
To compare ANRS$^{at}$ with ANRS$^{act}$, another observation is that category can help model to achieve a little improvement.  

\subsection{Hyperparameter Analysis}
In this section, we devise some experiments to explore the influence of two important hyperparameter in our model.
The one is the kernel size of CNN in the \textit{news aspect-level encoder}.
Another is the negative sample ratio $G$ in the model training procedure. 

\subsubsection*{Kernel Size of CNN}
\begin{figure}[htbp]
	\centering
	\includegraphics[width=6.5cm]{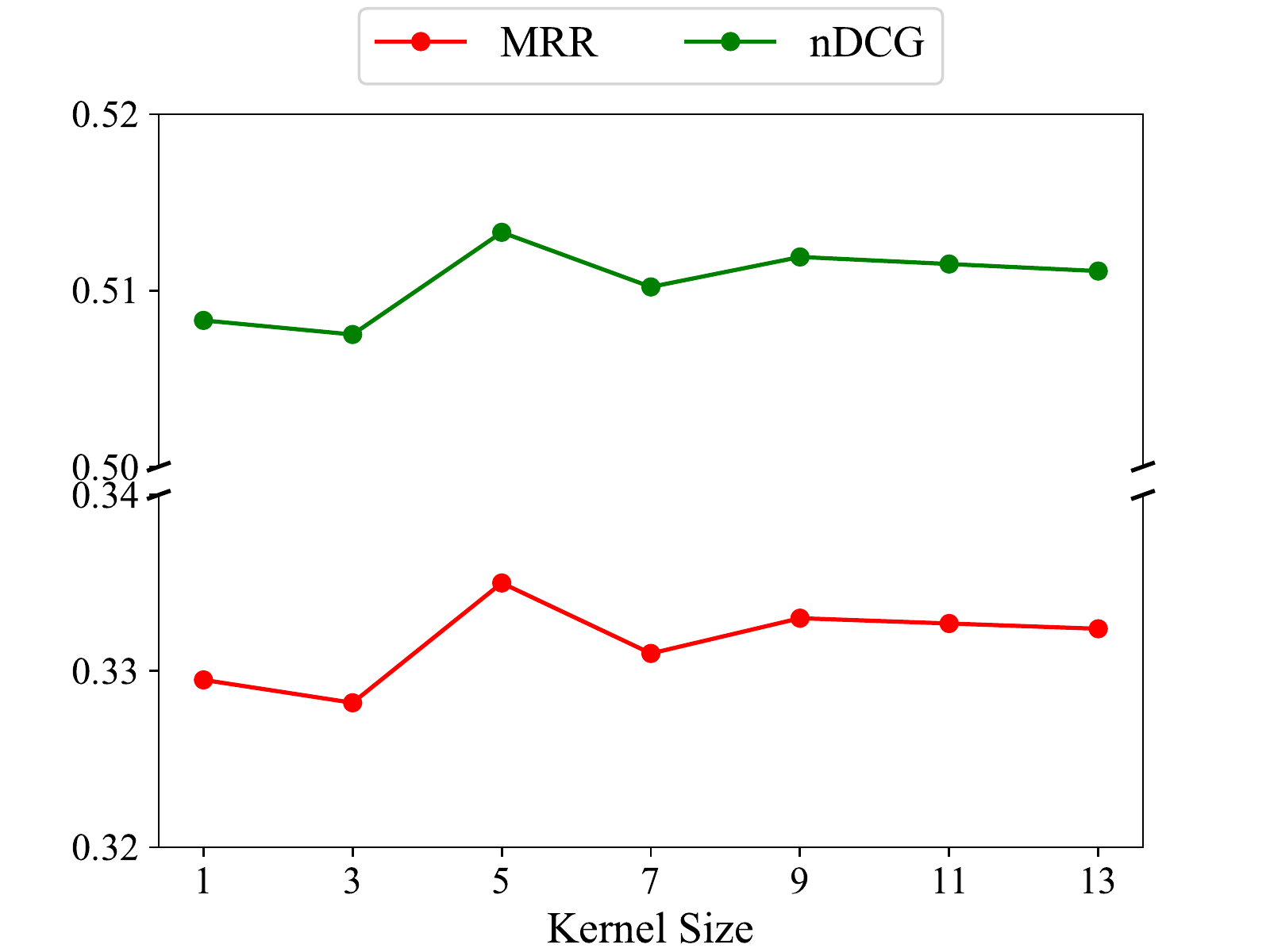}
	\caption{Performance of ANRS under different kernel size}
	\label{figcnn}
\end{figure}
We employ CNN to extract news features in the \textit{news aspect-level encoder}, so the kernel size of CNN will affect the performance of our model.
Aiming to find the optimal kernel size, we set it to 1, 3, 5, 7, 9, 11 and 13 respectively to verify the performance, as shown in Fig. \ref{figcnn}.
According to the figure, it is obvious that the model achieves the best performance when the kernel size is set to 5, which is adopted in our model.
This is probably because the small kernel size fails to capture long-distance context, while the large kernel size has a negative effect because of overfitting the noisy patterns.

\subsubsection*{Negative Sample Ratio}
\begin{figure}[htbp]
	\centering
	\includegraphics[width=6.5cm]{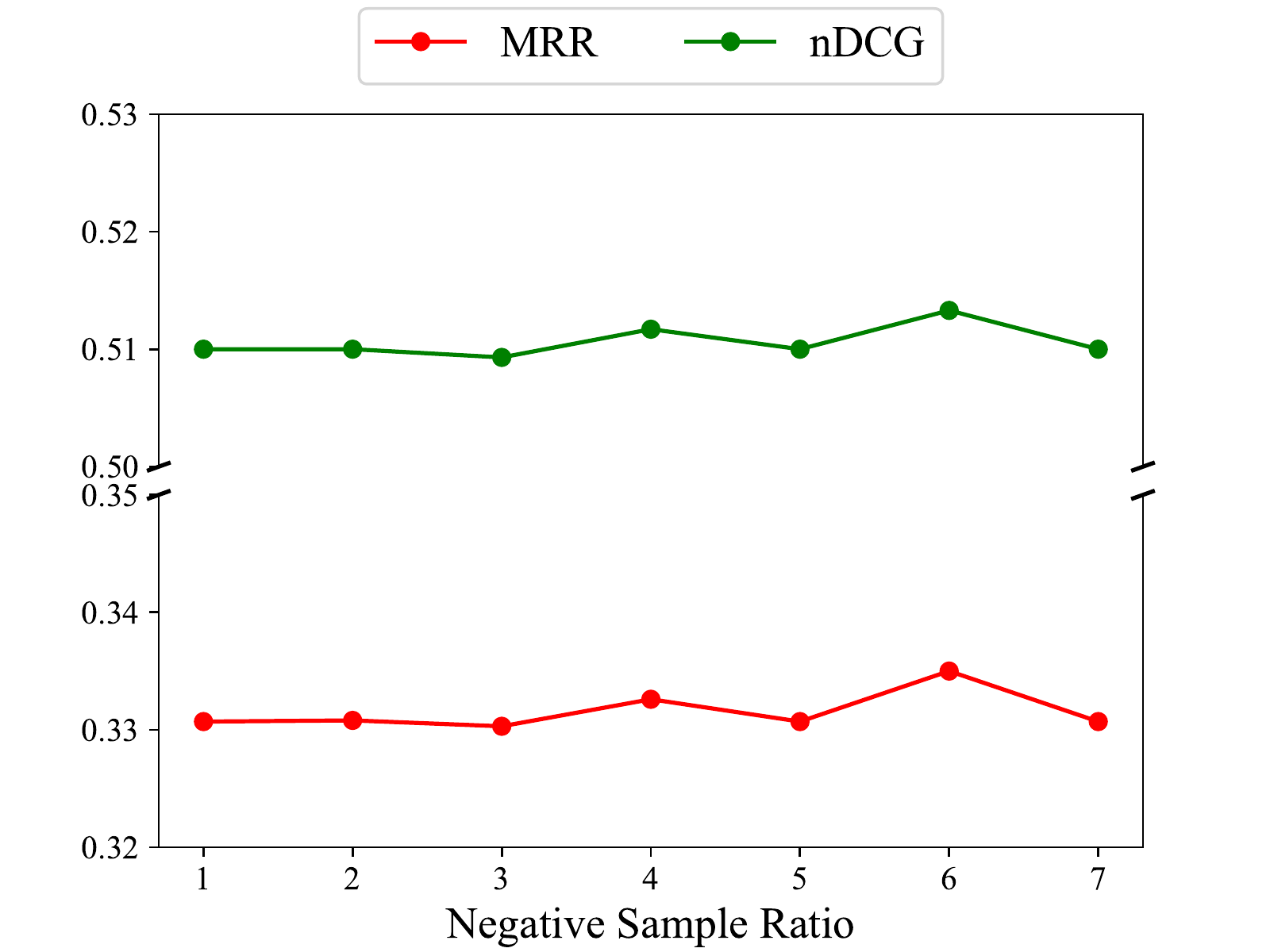}
	\caption{Performance of ANRS under different negative sample ratio}
	\label{K}
\end{figure}
During the training procedure, the negative sample ratio $G$ decides the number of negative samples.
Fig. \ref{K} illustrates the experimental results on different negative sample ratio $G$.
The performance reaches the peak when $G$ is set to 6. 
When $G$ is smaller than 6, the performance is not ideal. This is probably because that there is no enough negative samples to provide the sufficient information leading to unstable and sub-optimal performance.
When $G$ is larger than 6, the performance begins to drop down.
This is probably because that our model is difficult to identify the positive samples when $G$ is too large, leading to sub-optimal performance.
According to the figure, the best performance is achieved when $G$ is set to 6. 

\begin{figure*}[thpb]
	\centering
	\includegraphics[width=\textwidth]{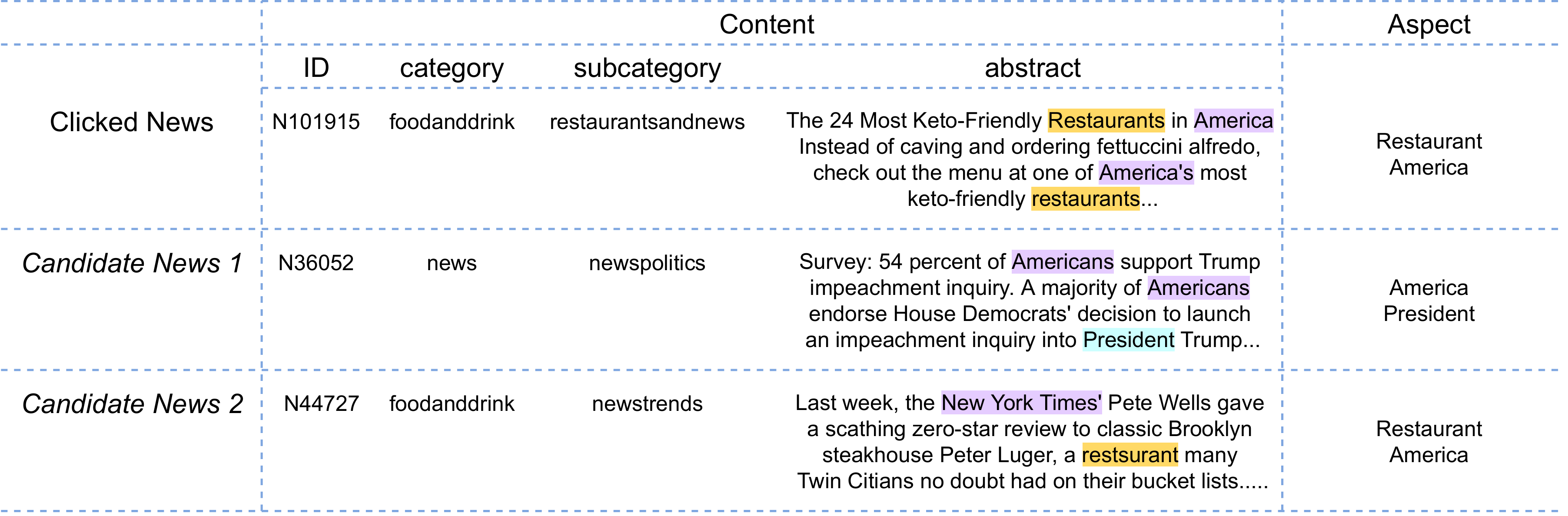}
	\caption{Case study of our model. 
		The news and users are randomly sampled from the dataset. The yellow, purple and blue highlights refer to the aspects \textit{Restaurant}, \textit{America} and \textit{President}.
	}
	\label{casestudy}
\end{figure*}
\subsection{Case Study}

To intuitively demonstrate the effectiveness of the aspect-level information, a case study is performed.
We randomly select the one clicked news including ID, category, subcategory and abstract in the test set.
As shown in Fig. \ref{casestudy}, we observe that our model can capture the important related words to learn the aspect information.
For example, because there are some representative words such as \textit{restaurant}, \textit{America} in the clicked news, although the category of this news is \textit{food and drink}, our model can learn different aspects, i.e., \textit{Restaurant}, \textit{America}.
Though there is not a location label in the category information of the clicked news, however, as its abstract mentions the location, i.e., \textit{America}, our model can capture and learn the location aspect automatically. 
In addition, our model can recommend some diversified news to users and satisfy their potential interests.
For example, the clicked news belongs to the category \textit{food and drink}  and the subcategory \textit{restaurants and news}.
The user who clicks this news may be interested in \textit{food} and \textit{keto-friendly restaurant}.
With our model, the user may be recommended with \textit{Candidate News 1}, which is related with the clicked news by the common aspect \textit{America}.
The user may enjoy a satisfied reading experience by the recommended diversified news.

\section{Conclusions}
In this paper, we propose a novel aspect-driven news recommender system (ANRS), which is built on aspect-level user preference and news representation learning. 
ANRS consists of three main modules, i.e., \textit{news aspect-level encoder}, \textit{user aspect-level encoder} and \textit{click predictor}. 
We utilize CNN and attention network to extract user's preferences and news features. Meanwhile, we extract aspect information to enhance the representation of users and news. 
Empirical evaluations on the real-world dataset demonstrate the superiority of our model.
In future work, we will explore the more effective neural architecture to accurately capture aspect information. 
Besides, the content of news may be more useful for modeling aspect features, which will be further explored.


%

\section*{Acknowledgment}
The research work is partly supported by National Key R\&D Program of China under Grant No.2018YFC0831700 and No.2018YFC0830705, National Natural Science Foundation of China under Grant No.61502259, and Key Program of Science and Technology of Shandong Province under Grant No.2020CXGC010901 and No.2019JZZY020124.

\ifCLASSOPTIONcaptionsoff
  \newpage
\fi



\bibliographystyle{IEEEtran}
\bibliography{ref}
%

%




\end{document}